\documentstyle[aas2pp4]{article}
\def \as   {$^{\prime\prime}$}  
\def \cmsq           {\hbox{cm$^{-2}$}}
\def \deg          {\ifmmode ^{\circ}\else $^\circ$\fi}  
\def \etal         {{\it et~al.} }
\def \flam         {\hbox{ergs s$^{-1}$ cm$^{-2}$~\AA $^{-1}$}}
\def \Ha           {\hbox{H$\alpha$}}

\def \kms          {\rm{\hbox{km s$^{-1}$}}}
\def \lam          {$\lambda$}
\def \Lya          {\hbox{Ly$\alpha$}}
\def \Lyb          {\hbox{Ly$\beta$}}
\def \pcc           {\hbox{cm$^{-3}$}}
\def \zaz          {{$z_a\kern -1.5pt \approx\kern -1.5pt z_e$}}
\def \zllz         {{$z_a\kern -3pt \lam\kern -3pt z_e$}}

\def \zgz          {{\kiA z\lower 3pt \hbox{a} $>$ z\lower 3pt \hbox{e}\ }}
\lefthead{Hamann et al.}
\righthead{Broad \protect\ion{Ne}{8}~\lam 774 Emission}

\begin{document}

\renewcommand{\baselinestretch}{1.5}

\title{\Large\bf Broad \protect\ion{Ne}{8}~\lam 774 Emission From Quasars}

\renewcommand{\baselinestretch}{1}

\author{FRED HAMANN\altaffilmark{1}, ROSS D. COHEN}
\affil{The Center for Astrophysics and Space Sciences, University of 
California -- San Diego, \\ La Jolla, CA 92093-0424}
\author{JOSEPH C. SHIELDS}
\affil{Department of Physics and Astronomy, Ohio University, Athens, 
OH 45701-2979}
\author{E. M. BURBIDGE, VESA JUNKKARINEN}
\affil{The Center for Astrophysics and Space Sciences, University of 
California -- San Diego, \\ La Jolla, CA 92093-0424}
\author{D. M. CRENSHAW} 
\affil{Goddard Space Flight Center, Code 681, Greenbelt, MD 20771}

\altaffiltext{1}{email: fhamann@ucsd.edu} 

\begin{abstract}

\normalsize
\ion{Ne}{8}~\lam 774 is an important tracer of the high-ionization 
gas in QSOs. We examine the \ion{Ne}{8} emission-line properties 
using new {\it HST}-FOS spectra of four sources, mean spectra derived 
from two QSO samples in the {\it HST} archives, and new 
photoionization calculations. 
The results support our previous claim that broad \ion{Ne}{8} 
lines are common in QSOs, with an average flux of $\sim$42\% of 
\ion{O}{6}~\lam 1034 and velocity widths that are 
$\sim$2 to 5 times larger  
than \ion{O}{6}, \ion{C}{4}~\lam 1549 and other broad lines in 
the same spectra. The strongest and most reliably measured  
\ion{Ne}{8}~\lam 774 lines (in two sources) have 
FWHM~$\sim$~14,500~\kms . Line profile fits in these cases show that 
the unusually large widths might be caused by blending with emission  
from \ion{N}{4}~\lam 765 and \ion{O}{4}~\lam 789. However, standard 
photoionization calculations indicate that \ion{N}{4}, \ion{O}{4} and 
all other lines near this wavelength should be too weak, leaving 
(very broad) \ion{Ne}{8} as the only viable identification for the 
$\sim$774~\AA\ feature. (This conclusion might be avoided if there are 
large radial velocity dispersions [$\ga$1000~\kms ] in the emitting region 
and the resonant absorption of continuum photons enhances the flux 
in weaker lines.) The calculations also indicate that the 
\ion{Ne}{8} emitting regions have ionization parameters in the range 
$5\la U \la 30$,  total hydrogen column densities of 
$10^{22}\la N_{\rm H} \la 3\times 10^{23}$~cm$^{-2}$, and 
an average covering factor of $\ga$30\% (for solar abundances 
and a nominal QSO continuum shape). The \ion{Ne}{8} emitting region is 
therefore more extensive, more highly ionized, and has much higher 
velocities than the rest of the broad emission line region (BELR). 
This highly-ionized BELR component would be a strong X-ray ``warm'' absorber 
if it lies along our line-of-sight to the X-ray continuum source. 

\end{abstract}

\keywords{Galaxies: Quasars: emission lines -- Galaxies: Quasars: general}

\section{Introduction}

The rest-frame UV spectra of QSOs and active galactic nuclei (AGNs) 
typically contain strong and broad emission lines (BELs) from the 
resonance doublets \ion{C}{4}~\lam 1549, \ion{N}{5}~\lam 1240 and 
\ion{O}{6}~\lam 1034 in the lithium-like isoelectronic sequence. 
\ion{Ne}{8}~$\lambda$774 is another resonant doublet in that sequence. 
Its short wavelength makes \lam 774 difficult to measure, 
but this line deserves special attention because \ion{Ne}{8} 
requires ionization energies (207~eV) nearly 
twice as large as \ion{O}{6} (114~eV) and more than four times larger 
than \ion{C}{4} (48~eV). Broad \ion{Ne}{8}~$\lambda$774 lines 
can therefore probe the extreme 
high-ionization gas in the broad emission line regions (BELRs) and 
test the possible relationship of this gas to other 
high-ionization phenomena in QSO/AGN environments. 
For example, the \ion{Ne}{8}-emitting BELR has an ionization 
consistent with the X-ray ``warm'' absorbers (Halpern 1984) 
observed in some quasars and AGNs. Warm absorbers can be identified 
by bound-free edges of \ion{O}{7} and \ion{O}{8} near 0.8~keV 
(for example, 
\cite{geo97,rey97,ota96,mih94,fab94,tur93a}). Photoionization calculations 
indicate that the regions producing \ion{Ne}{8} line emission would 
also produce these edges in soft X-rays if the gas lies along our 
line-of-sight to the X-ray continuum source 
(\cite{net93,ham95a,shi95,rey95,net96}). 

There are now several reports of broad \lam 774 emission from QSOs 
(\cite{coh94,ham95a,zhe97}, Hamann, Zuo \& Tytler 1995b), but the 
data are often of poor quality and the number of secure detections is 
small. Hamann \etal (1995b) made the first attempt to 
study the general strength and character of 
\ion{Ne}{8}~\lam 774 emission in a sample of QSOs.  
That sample consists of 5 radio-loud and radio-quiet sources 
measured in the {\it Hubble Space Telescope} ({\it HST}) Absorption Line 
Snapshot Survey (Absnap). They found that 3 of the 5 sources,
and possibly all 5, have a BEL at $\sim$774~\AA . 
They argued that the measured wavelengths and  
fluxes (compared to photoionization models) both point to
\ion{Ne}{8}~\lam 774 as the most likely identification. 
More recently, Zheng \etal (1997) measured the $\sim$774~\AA\ emission 
line in a composite spectrum derived from a larger number of QSOs. 

Here we supplement that work with new photoionization 
calculations and new {\it HST} and ground-based spectroscopy of 
four QSOs. We also extract an unbiased sample of 11 QSOs from the 
{\it HST} archives and combine them with our previous 
Absnap sample to construct mean spectra for a total of 16 sources. 
The new individual spectra provide the highest signal-to-noise 
measurements of the \ion{Ne}{8} feature, while the means allow us to 
study the \ion{Ne}{8}~\lam 774 emission properties in representative 
samples. Section 2 below describes the observations and measurements of 
the emission lines. Section 3 reexamines the \ion{Ne}{8}~\lam 774 
identification and provides an analysis of the kinematics 
and physical conditions 
based on photoionization simulations and fits to the line profiles. 
Section 4 discusses the implications of the \ion{Ne}{8}~\lam 774 
emission and, in particular, the possible relationship to warm absorbers. 
Section 5 provides a summary. 

\section{The Data}

\subsection{Observations and Target Selection}

We obtained UV spectra of four bright QSOs with the Faint Object 
Spectrograph (FOS) on board {\it HST}. We observed 
PKS~0355$-$483 and Q~1435$-$015 because they were 
strong candidates for broad \ion{Ne}{8}~\lam 774 emission 
in our previous study of the {\it HST}-Absnap sample (\cite{ham95b}). We 
chose PG~1148+549 as another candidate for strong \ion{Ne}{8} emission 
based on earlier observations with the {\it International Ultraviolet 
Explorer} ({\it IUE}; \cite{ham95a}). We observed the QSO 
PG~1522+101 as part of an unrelated absorption-line study 
and include it here because the {\it HST} spectra contain a strong 
emission line near 774~\AA\ rest (\cite{coh94}). 

The {\it HST} spectra encompass 
\ion{Ne}{8}~\lam 774 and other prominent lines in the rest frame 
UV but, for the two highest-redshift sources, Q~1435$-$015 and 
PG~1522+101, they do not include the important \ion{C}{4}~\lam 1549 
line. We therefore obtained ground-based optical spectra of these 
QSOs with the Shane 3.0 m telescope at Lick 
Observatory. Table 1 provides a log of the observations, including the 
dates, the telescopes used (Tel.), the spectrographs and their 
setups/gratings (Instr./Setup), the wavelengths covered ($\lambda_{obs}$), 
the approximate full-width-at-half-maximum resolutions 
$R\equiv \lambda/\Delta\lambda$, 
and the total exposure times (Exp.). The emission-line 
redshifts listed with the QSO names are from Hewitt \& Burbidge (1993), 
except for PG~1148+549, which is from Schmidt \& Green (1983). 
The {\it HST}-FOS spectra were obtained through the 1\as\ science 
aperture and calibrated at the 
{\it Space Telescope Science Institute} using their standard ``pipeline'' 
procedures. We recalibrated the absolute wavelengths in the G130H and G270H 
spectra by requiring that Galactic absorption lines appear at their 
laboratory wavelengths. We then shifted the G190H or G160L spectra by  
matching features in spectral regions that overlap with G130H or 
G270H. We expect that the final $3\sigma$ wavelength uncertainties are 
less than 0.5 diodes, that is $\la$4~\AA\ with the G160L 
grating and $\la$0.5--1~\AA\ with the higher resolution gratings. 
The Lick observations 
employed a 2\as\ slit and a $400\times 1200$ Reticon CCD. We reduced 
those data in the usual way using VISTA software (modified 
slightly by Dr. T. A. Barlow). The wavelength uncertainties in the Lick 
spectra should be less than $\sim$0.2~\AA . We flux-calibrated those 
spectra by observing both the QSOs and standard stars through a wide slit 
on the same night. None of the data are corrected for Galactic extinction. 
We simply note that for a typical high-latitude Galactic column density 
of $N_H = 3\times 10^{20}$~\cmsq\ 
(Murphy \etal 1995 and references therein), a conversion factor of 
$E(B-V) = N_H/(4.93\times 10^{21}$~\cmsq ) (\cite{dip94}) and 
a typical extinction curve with $R_V = 3.1$ (Cardelli, Clayton \& Mathis 
1989), the extinction peak at the $\sim$2175~\AA\ 
silicate feature is roughly 0.5$^m$. This peak coincides 
with the \ion{O}{6}~\lam 1034 emission line for $z_e = 1.10$. 

Figures 1--4 plot the new and old spectra together. 
The new {\it HST} observations of 
PKS~0355$-$483 and Q~1435$-$015 (Figures 1 and 3) have higher 
signal-to-noise ratios and slightly higher resolutions than 
the Absnap data because of the longer exposure times and improved 
line spread function of the post-COSTAR FOS. 
Figure 2 includes two versions 
of the {\it IUE} spectrum of PG~1148+549 for comparison to the 
{\it HST} data. The bottom spectrum in that figure, labeled ``old {\it IUE}'', 
is the original data as shown in Hamann \etal (1995a). The middle 
spectrum, labeled ``new {\it IUE}'', is a new reduction of the same 
data. The old {\it IUE} spectrum was reduced using an optimal 
extraction technique (\cite{kin91,lan93}). The new {\it IUE} 
spectrum derives from the  improved image processing system 
(NEWSIPS) used for the {\it IUE} Final Archives (\cite{nic96}) and 
is the average from three separate images that have been cleaned of
cosmic rays. The new {\it IUE} reduction of PG~1148+549 has a 
slightly higher  signal-to-noise ratio, but there is a camera artifact 
at $\sim$1663~\AA\ (Crenshaw \etal 1990) that is enhanced 
by NEWSIPS. The \ion{Ne}{8} emission line discussed by 
Hamann \etal (1995a) is weaker in both the NEWSIPS 
reduction and the {\it HST} spectrum 
(upper curve in Figure 3).  

Figure 4 plots the low (G160L) and high (G190H+G270H) 
resolution {\it HST} spectra of PG~1522+101, with the longer 
wavelength Lick Observatory data appended to the latter. 
Both {\it HST} observations of this source show 
strong $\sim$774~\AA\ emission, but the higher resolution spectrum shows 
more clearly the ``forest'' of \Lya\ absorption lines shortward of the 
QSO's \Lya\ emission. These absorption features can 
significantly distort the continuum and BELs in low resolution data. 
There is also weak Lyman limit absorption in this spectrum at 
$\sim$1980~\AA\ (observed), which is confirmed by detection of the 
corresponding Lyman series lines. 

\subsection{Mean Spectra}

We constructed mean spectra of two samples of QSOs measured 
with {\it HST}-FOS and the low-resolution grating (G160L). 
The first sample consists of the 5 QSOs selected by Hamann \etal 
(1995b) from the {\it HST}-Absnap database (\cite{tyt94}). 
That sample includes QSOs with 
(1) emission redshifts ($z_{e}\approx 0.75$ to 1.4) that allow 
wavelength coverage across both \ion{Ne}{8}~\lam 774 
and \ion{O}{6}~\lam 1034 with the G160L grating, 
(2) no Lyman-limit absorption across $\sim$774~\AA\ in the QSO rest 
frame,  and (3) subjectively ``measurable'' \ion{O}{6} emission. 
The last criterion serves as a signal-to-noise 
discriminator and insures that we can establish an emission redshift 
for coadding the spectra. 
This selection based on \ion{O}{6} could introduce a bias toward 
sources with more high-ionization gas and thus stronger \ion{O}{6} and 
\ion{Ne}{8} emission lines (see \S2.4 below). However, 
the QSOs were {\it not} chosen for their \ion{Ne}{8} characteristics.

The second sample consists of 11 QSOs from the first {\it HST} Absorption 
Line Key Project database (\cite{bah93}) selected by criteria (1) and (2) 
above\footnote{We excluded a 12th object 
from the Key Project sample, PG~1407+265, even though it meets both criteria 
1 and 2. This peculiar object has extremely weak lines (we estimate 
REW(\ion{O}{6}) and REW(\Lya )~$<$6~\AA\ at 3$\sigma$) and a poorly 
defined redshift (see \cite{mcd95}).}. 
Seven of these QSOs have measurable \ion{O}{6} 
emission and 4 do not. We will refer to the subset of 
7 Key Project spectra with significant \ion{O}{6} 
as KP-sub, and the full sample of 11 Key Project spectra as 
KP-full\footnote{The KP-sub sample is the same as what we called ``Key 
Project'' in our preliminary report on these data (\cite{ham96b}), 
except here we have identified and included one more QSO that meets 
the sample criteria.}. 
Note that none of the spectra that lack \ion{O}{6} emission, including 
PG~1407+265 (footnote 2), have a detectable \ion{Ne}{8} line. 

Figure 5 plots the mean spectra derived from the various 
sub-samples and the Total sample of 16 sources. 
We calculated the means using the new {\it HST}-FOS spectra of the 
two Absnap sources PKS~0355$-$483 and Q~1435$-$015 (\S2.1). 
The other Absnap spectra are from Hamann \etal (1995b). 
The Key Project spectra are from the {\it HST} archives with reductions 
by Schneider \etal (1993). PG~1148+549 and PG~1522+101 (\S2.1)
are not in the Absnap or Key Project samples and so are not included in 
the means. For the 12 sources with 
well-defined \ion{O}{6} lines (in the Absnap and KP-sub samples) 
we shifted the individual spectra to the rest frame by forcing the 
centroid of the upper $\sim$50\% of the \ion{O}{6} profile to be at 
1033.8~\AA . For the 4 
sources without \ion{O}{6}, we shifted the spectra using published 
redshifts in Hewitt \& Burbidge (1993). Applying the published redshifts, 
which derive from low-ionization lines such as \ion{C}{3}]~\lam 1909 
or \ion{Mg}{2}~\lam 2799, 
leads to larger uncertainties (by at least several \AA ) in the 
rest wavelengths of \ion{O}{6} and \ion{Ne}{8} because 
(1) these redshifts do not compensate for the calibration uncertainties in 
the {\it HST}-FOS wavelengths and (2) there are known intrinsic redshift 
differences between high- and low-ionization lines (\cite{esp89,tyt92}).  
In all cases, we normalized the continua to unity by fitting\footnote{All 
fitting and arithmetic manipulations were performed with 
the IRAF software distributed by the National Optical Astronomy 
Observatories under contract to the National Science Foundation.} 
with low-order polynomials and then averaged the spectra with equal 
weights. Wavelength regions at the  
ends of individual spectra were excluded from the 
means if they contain obvious Lyman-limit absorption, 
severe noise spikes, or geocoronal emission lines such as \Lya\ 
in first or second order. The histogram at the bottom of Figure 5 
indicates the number of spectra contributing to the Total mean at 
each wavelength. Median spectra drawn from these samples 
are qualitatively the same as the means.  

\subsection{Line Measurements} 

Table 2 lists the centroid rest wavelengths ($\lambda_{rest}$), 
rest equivalent widths (REWs), observed fluxes (Flux), and full 
widths at half maximum (FWHM)  
for emission lines detected in the new {\it HST} and Lick spectra  
(Figures 1--4) and in several of the mean spectra (Figure 5). 
Not all of the lines labeled in Figures 1--5 are present in the data;  
for example, \ion{Mg}{10}~\lam 615 and \ion{Ne}{7}]~\lam 895 
are not detected in any source with approximate 3$\sigma$ upper 
limits of REW~$\la 2$~\AA\ in all cases. The 
measurements for PG~1522+101 are from the higher resolution data after 
interpolating across the strongest of the overlying \Lya\ forest lines. 
We made no attempt to correct for absorption lines in the 
other sources. The fluxes given 
for the mean spectra are relative to \ion{O}{6}~\lam 1034, while 
those given for the individual sources are as observed 
with units 10$^{-14}$ ergs s$^{-1}$ cm$^{-2}$. Recall that the 
data are not corrected for Galactic extinction (\S2.1). The 
equivalent widths are immune to extinction effects, but the fluxes 
are not. In a worst-case scenario, with the $\sim$2175~\AA\ extinction 
peak overlying the \ion{O}{6} line (\S2.1), the differential extinction 
between \ion{O}{6} and \ion{Ne}{8} for a typical Galactic column 
density would be $\sim$0.1$^m$ and would 
lead us to overestimate the true \ion{Ne}{8}/\ion{O}{6} flux ratio by 10\%.

The emission lines were measured by integrating 
the flux above the fits to the continua (dotted curves in Figures 1--5) 
between the wavelengths indicated 
by $\Delta\lambda$ in Table 2. We did not attempt to separate 
severely blended lines or line components. In particular, 
the feature we attribute to 
\ion{O}{6}~\lam 1034 could have a contribution from \Lyb\ \lam 1026. 
Laor \etal (1995) and Wills \etal (1995) estimate that the mean \Lyb\ 
contributions to this blend 
are $\sim$23\% and $\la$10\%, respectively, in samples of 
low-redshift QSOs measured with {\it HST}. We will ignore 
this small contribution in our discussions below. We will also 
ignore the weak \ion{Si}{2}~\lam 1527 line that can add to the measured 
\ion{C}{4}~\lam 1549 flux.

The feature we identify as \ion{Ne}{8}~\lam 774 is clearly present in 
the mean spectra, and it is 
stronger than the means (both in REW and relative to other 
lines) in the individual QSOs PKS~0355$-$483 (radio-loud) and 
PG~1522+101 (radio-quiet). The observations of Q~1435$-$015 are 
less conclusive, but the new data are consistent with the previous 
report of strong $\sim$774~\AA\ emission from this source (\cite{ham95b}). 
The new {\it HST}-FOS spectrum of PG~1148+549 does not exhibit 
the strong $\sim$774~\AA\ line noted in the original {\it IUE} data 
(\cite{ham95a}). The {\it HST} spectrum of this source resembles 
the reprocessed (``new'') {\it IUE} spectrum (Fig. 3), indicating that the 
\ion{Ne}{8} line is weaker than our previous claim. 

There is evidence in the means and in some of the 
individual spectra for emission from the excited-state line 
\ion{C}{3}$^*$~\lam 1176 and from a broad unidentified feature at 
$\sim$1070~\AA . The unidentified feature was noted previously 
by Laor \etal (1994 and 1995) in their sample of low-redshift QSOs. 
The \ion{C}{3}$^*$ line could prove to be a valuable 
diagnostic of BELR physical conditions because it involves a transition 
between unusually 
high energy states, 17.1 to 6.5 eV. The lower state is metastable 
and gives rise to the well-known \ion{C}{3}]~\lam 1909 emission line 
(see \cite{kor97,lao97a} for discussion). 

\subsection{What is the Average $\sim$774~\AA\ Line Strength?}

Unbiased measurements of the average $\sim$774~\AA\ line strength 
require spectra of many QSOs selected by only 
their redshifts and their lack of overlying Lyman-limit absorption 
(criteria 1 and 2 in \S2.2). Among the data we collected, the KP-full 
sample of 11 spectra comes closest to this ideal. The average $\sim$774~\AA\ 
line in that sample has REW~$\sim$~5.9~\AA\ and a flux of $\sim$42\% of 
\ion{O}{6}~\lam 1034 (Table 2). These results can be compared to 
REW~$\sim$~3.2~\AA\ and a flux ratio of \lam 774/\lam 1034~$\sim$~21\% 
reported by Zheng \etal (1997) from their composite {\it HST} spectrum.  
The Zheng \etal composite includes 
$\sim$20 spectra across $\sim$774~\AA\ and 80--90 across the 
\ion{O}{6} line. Our KP-full sample is undoubtedly a subset of the 
spectra they used. The different line strengths in these 
two samples could be caused by random fluctuations in the small numbers 
of spectra involved and/or by the large measurement uncertainties, which are 
dominated by the subjective continuum placement. However, the reported 
differences might also be caused by selection effects. In particular, 
there is an inherent ``Baldwin Effect'' in the Zheng \etal 
composite. Several studies have shown that this effect is pervasive in 
QSO samples; more luminous QSOs have 
typically lower emission-line REWs (\cite{bal77,kin90}). The spectra in 
the {\it HST} database are primarily of the brightest QSOs at 
each redshift. Our KP-full sample includes a relatively narrow redshift 
range so that $\sim$774~\AA\ through $\sim$1034~\AA\ are measured in the 
same data, but the Zheng \etal composite has many more 
low redshift and low luminousity QSOs contributing at long wavelengths 
(e.g. across \ion{O}{6}) than at short wavelengths (at \ion{Ne}{8}). 
Therefore their composite should have a larger REW in \lam 1034, 
a smaller REW in \lam 774 and a lower \lam 774/\lam 1034 strength 
ratio compared to our KP-full sample, which agrees with the measured 
results.  

Therefore, our KP-full sample, although still limited by the small number 
of spectra and significant measurement uncertainties, should 
provide the most representative measure 
of the \lam 774 line strength and the \lam 774/\lam 1034 flux ratio in 
(bright) QSOs at redshift $\sim$1. The Absnap and KP-sub samples 
both have stronger \lam 774 lines than KP-full, probably because of a 
selection bias introduced by the requirement for ``measurable'' 
\ion{O}{6} emission (\S2.2). Note, however, that while \lam 774 
is stronger in the Absnap sample, \lam 1034 and the \Lya\ blend are not; 
therefore, the nature and extent of the selection bias is ambiguous.  
The Total mean mixes the Absnap and KP-full samples and therefore 
has slightly stronger $\sim$774~\AA\ line emission than KP-full alone. 

\section{Analysis}

\subsection{Photoionization Calculations}

Theoretical models of the BELR are needed 
to understand the conditions under which \ion{Ne}{8}~\lam 774 
forms and the likelihood that other lines near this wavelength might 
contribute to the measured emission. Some recent 
photoionization calculations have made specific predictions 
for the \ion{Ne}{8} line strength  
(\cite{ham95a,kor97,net96}). Here we update the 
calculations in Hamann \etal (1995a) using a newer version of 
CLOUDY (90.02; \cite{fer96}) to examine other strong lines at 
wavelengths $<$1000~\AA\ and test whether recent improvements in the 
atomic data affect the results (see also \cite{bal96,kor97}). 

\subsubsection{Predicted Line Strengths and the \protect\ion{Ne}{8} 
Identification} 

The most likely contributors to broad line emission near 774~\AA\ 
are \ion{Ne}{8}~\lam 774, \ion{N}{4}~\lam 765, \ion{O}{4}~\lam 789, 
\ion{S}{5}~\lam 786 and \ion{Ar}{6}~\lam 763 (Verner, Verner \& Ferland 
1996). The calculations by Hamann \etal (1995a) indicate that the 
alternatives to \ion{Ne}{8} should be relatively weak, and 
therefore \ion{Ne}{8} is the most likely identification for the measured 
line. Figure 6 shows the line equivalent widths predicted by 
CLOUDY for different values 
of the ionization parameter $U$ (defined as the dimensionless ratio 
of hydrogen-ionizing photon to hydrogen particle densities at the 
illuminated face of the clouds). This plot is analogous to Figure 
2 in Hamann \etal (1995a). 
The model clouds have a space density of 
$n_H = 10^{11}$~\pcc , a column density of 
$N_H = 3\times 10^{23}$~\cmsq\ and solar element abundances. They 
are illuminated on one side by a standard QSO spectrum 
that is believed to be typical of low-redshift QSOs and Seyfert 1 
nuclei (\cite{mat87}). We modified 
this spectrum to have a sharp decline at 
wavelengths longer than 2~$\mu$m to avoid significant free-free heating 
(\cite{fer92}). The internal cloud velocities are assumed to 
be thermal. The equivalent widths in Figure 6 
apply for emitting regions that completely cover the central 
QSO (over 4$\pi$ ster) but are transparent to the   
radiation from other clouds (e.g. \Lya\ and \ion{Ne}{8} 
photons emitted on the far side of the QSO are not absorbed or 
reflected by clouds on the near side). 

The effects of different 
space densities and spectral shapes on some of the line strengths 
are shown by Korista \etal (1997) and Netzer (1996). In general, the 
relative strengths of the permitted lines in Figure 6 are only 
weakly dependent on the space density for a given $U$. The intercombination 
lines \ion{O}{5}] and \ion{Ne}{7}] are strongly suppressed above 
their common critical density of $\sim$$10^{11}$~\cmsq . 
Smaller column densities would significantly 
reduce the emission from low- to intermediate-ionization 
lines (e.g. \Lya , \ion{C}{4} and \ion{N}{5}) at the right in Figure 6, 
because these lines form strictly at large depths in the high-$U$ clouds 
(see Fig. 3 in Hamann \etal 1995a). 
Examination of the heating and cooling processes indicates that, 
for the continuum shape used here, clouds with 
$U\ga 30$ are thermally unstable (\cite{ham95a,net96}). 

The only important difference between Figure 6 here and Figure 2 in 
Hamann \etal (1995a) is that the 
\ion{O}{4} line is now several times stronger, particularly at low and 
intermediate $U$; but the main results 
are the same. \ion{S}{5}~\lam 786 and \ion{Ar}{6}~\lam 763 
are too weak under any circumstances to produce the measured 
$\sim$774~\AA\ line. (The predicted \ion{Ar}{6}~\lam 763 line, not 
shown in Fig. 6, is at least 5 times weaker than \ion{S}{5}~\lam 786.) 
The \ion{N}{4}~\lam 765 and \ion{O}{4}~\lam 789 lines 
are the strongest alternatives to \ion{Ne}{8}~\lam 774. They form in 
roughly the same gas as \ion{C}{4}~\lam 1549, so their strengths 
relative to \ion{C}{4} are not sensitive to the uncertain 
geometry or physical conditions. 
Scaling the predicted \ion{N}{4}/\ion{C}{4} and 
\ion{O}{4}/\ion{C}{4} ratios in Figure 6 by the \ion{C}{4} 
measurements in Table 2 shows that the observed 774~\AA\ lines 
in PKS~0355$-$483 and PG~1522+101 are much too strong 
to have significant contributions from \ion{N}{4} or 
\ion{O}{4}. In particular, for $U\la 1$ (see \S4), \ion{N}{4}+\ion{O}{4} 
should have REW~$\la$~0.6~\AA\ in PKS~0355$-$483 and REW~$\la$~1.9~\AA\ 
in PG~1522+101, compared to the measured value of REW~$\sim$~13.5~\AA\ 
in both QSOs. The ratio of the fluxes predicted in \ion{N}{4}+\ion{O}{4} 
to that measured in $\sim$774~\AA\ is therefore $\la$0.07 in PKS~0355$-$483 
and $\la$0.3 in  PG~1522+101. 

The mean spectra (Fig. 5) 
do not include \ion{C}{4}~\lam 1549 for comparison, 
but in large ground-based samples 
\ion{C}{4} has REW~$\sim$~20 to 32~\AA , or about $\sim$50\% of 
\Lya\ (see the compilation by \cite{ham96}). If we adopt a 
conservatively large value of REW(\lam 1549)~=~40~\AA , then Figure 6 
implies that the average \ion{N}{4}+\ion{O}{4} emission should have 
REW~$\la 1.0$~\AA\ for $U\la 0.1$ or REW~$\la 3.2$~\AA\ for $U\la 1$. 
These estimates are no more than about half the measured mean of 
REW~$\sim$~5.9~\AA\ in the $\sim$774~\AA\ line (\S2.4). 
Since values of $U\la 0.1$ are generally favored for the 
\ion{C}{4} emitting region (\S4), the average contributions of 
\ion{N}{4} and \ion{O}{4} to the $\sim$774~\AA\ feature should be 
$\la$17\%.

These results indicate that \ion{Ne}{8}~\lam 774 dominates the 
average $\sim$774~\AA\ emission and is the only significant 
contributor whenever this line is strong compared to \ion{C}{4}. 
Experiments with CLOUDY show that this conclusion holds even if the 
metallicities are above solar and nitrogen is several times 
overabundant relative to the other metals (\cite{ham92} and 1993).  

\subsubsection{Ionization, Column Densities and Covering Factors}

Figure 6 also supports the conclusions in Hamann \etal (1995a) regarding 
the ionization, column densities and covering factors of the emitting 
regions. Significant emission in \ion{Ne}{8}~\lam 774, relative to both the 
continuum and other lines, requires ionization parameters $U\ga 5$. 
Thermal stability in the clouds independently requires $U\la 30$, 
for an overall range of $5\la U\la 30$ in the \ion{Ne}{8} line-forming gas. 
The ratio of the observed mean (5.9~\AA ; \S2.4) 
to largest-predicted ($\sim$20~\AA ; Fig. 6) \lam 774 equivalent widths 
implies an average minimum covering factor of $\sim$30\%. The 
\ion{Ne}{8} emitting regions must subtend at least this fraction of the 
sky as seen from the central QSO to produce the average equivalent 
width. (The same analysis shows that the 
strongest \ion{Ne}{8} emitters, PKS~0355$-$483 and 
PG~1522+101, require covering factors $\ga$65\%.) 
The observed mean \lam 774 equivalent width also requires a 
minimum hydrogen column density of $N_H\ga 10^{22}$~\cmsq\ for 
solar abundances, even if the emitting region completely covers the 
central QSO (see Figure 3 in \cite{ham95a}). (The strongest \ion{Ne}{8} 
emitters require column densities at least twice as high.) 
For the range in $U$ values 
given above, column densities beyond $3\times 10^{23}$~\cmsq\ do not add 
to the \ion{Ne}{8} flux, so we have an effective upper limit of 
$N_H\la 3\times 10^{23}$~\cmsq\ for \ion{Ne}{8} emission.

These results are uncertain by factors of a few because they depend 
on the poorly known shape of the ionizing spectrum. 
The ionization parameter 
reflects the integrated continuum flux at energies $\ga$13.6~eV, 
but the \ion{Ne}{8} ionization depends only on the flux above 207~eV. 
Therefore, different continuum shapes will yield different estimates of 
the minimum $U$ value. Netzer (1996) argued that highly ionized species  
like \ion{Ne}{8} are best described by an X-ray ionization parameter, 
$U_x$, that includes energies between only 0.1 and 10 keV. 
A similar problem is that the equivalent widths measure the 
line strengths relative to the {\it local} continuum (e.g. at 
774~\AA ), even though that continuum 
has little to do with the production of \ion{Ne}{8}~\lam 774. 
Recent calculations by Netzer (1996) predict $\sim$3 
times lower equivalent widths for \lam 774 compared 
to Figure 6 because of a softer assumed continuum; the flux 
ratio $F_{\nu}(207{\rm eV})/F_{\nu}(774{\rm~\AA})$ 
is $\sim$3 times smaller in Netzer's (1996) continuum than in 
the Mathews \& Ferland (1987) continuum used in Figure 6. 
Conversely, experiments with CLOUDY show that harder (flatter) continua, 
such as the single power law $F_{\nu}\propto\nu^{-1.4}$ or the 
``baseline'' continuum used by Korista \etal (1997), roughly 
double the peak \ion{Ne}{8} equivalent width compared to 
Figure 6 and shift the REW curve to more than 5 times lower $U$. 
Even harder power-law continua, such as  
$F_{\nu}\propto\nu^{-1.2}$, still only double the peak \ion{Ne}{8} 
equivalent width. The ratio of observed ($\sim$5.9~\AA ) to 
largest-predicted ($\sim$40~\AA ) equivalent widths in these extreme  
cases therefore implies a lower limit on the average \ion{Ne}{8} covering 
factor of $\sim$15\%. 

\subsubsection{A Caveat: Continuum Pumping in High-Velocity Environments}

Another uncertainty is the importance of continuum pumping to the line 
emission. If there are large velocity dispersions along radial 
lines of sight through the BELR, the resonant absorption 
of continuum photons can ``pump'' electrons into upper energy 
states and significantly enhance the emission line fluxes. 
This pumping can take two forms, (1) resonant 
line scattering (absorption and re-emission in the same transition), 
and (2) continuum fluorescence (absorption into high energy states 
followed by cascades through intermediate energy levels). These 
processes are negligible for metallic lines in ``standard'' BELR models 
(e.g. Fig. 6 above) because the continuum photons encounter only thermal 
line widths in the BELR. However, if there is substantial turbulence 
or a large range of radial velocities (e.g. in an outflow), 
the absorption of continuum photons 
can compete with other excitation mechanisms -- particularly in weak 
resonance lines. Since the velocity field in the BELR is essentially 
unknown, we consider the possible effects of line scattering and 
continuum fluorescence here. 

The amount of scattering in a given line depends on its velocity width and 
optical depth as encountered by the continuum radiation. 
For unblended resonance lines that (1) have no alternate decay routes, 
(2) are not collisionally deexcited (e.g. at high densities), and (3)
are on the flat part of the curve-of-growth (optically thick but no 
damping wings), the scattering equivalent width resulting from  
thermal or micro-turbulent gas velocities can be derived by 
integrating over the saturated absorption line profile. 
This integration yields 
\begin{equation}
W_{\lambda}^{scat}\ \approx\ 2 f_c \lambda_o 
\left({{V_D}\over{c}}\right)\sqrt{\ln\left({{\tau_o}\over{\ln 2}}\right)}\ \ \ ,
\end{equation}
where $f_c$ is the covering factor of the emission region 
($0\leq f_c\leq 1$), $\lambda_o$ is the line wavelength, $\tau_o$ is the 
line-center optical depth, and $V_D$ is the Doppler velocity (thermal 
or turbulent). Similarly, radial outflows  
spanning a velocity range $\Delta V_r$ in optically thick lines 
will produce scattering equivalent widths of  
\begin{equation}
W_{\lambda}^{scat}\ \approx\ f_c \lambda_o \left({{\Delta V_r}\over{c}}\right)
\ \ \ .
\end{equation}
Keeping in mind that $\tau_o\propto V_D^{-1}$ in Eqn. 1 and that 
well-separated doublets like \ion{Ne}{8}~\lam\lam 770,780 scatter twice 
as much flux as single or blended lines, one can use the values of 
$\tau_o$ plotted by Hamann \etal (1995a) for strictly thermal velocities 
to estimate $W_{\lambda}^{scat}$ for different values of $V_D$. 
Here we use CLOUDY to derive the $\tau_o$ directly for models 
identical to Figure 6 but with larger $V_D$. We find 
that for clouds with $U=10$ and $V_D = 1000$~\kms , 
the \ion{Ne}{8} doublet has $W_{\lambda}^{scat} \sim 25 f_c$~\AA . 
Similarly, for $U=0.1$ and $V_D = 1000$~\kms , \ion{C}{4}~\lam 1549 
has $W_{\lambda}^{scat} \sim 30 f_c$~\AA\ and the \ion{N}{4}~\lam 765 
and \ion{O}{4}~\lam 789 lines both have 
$W_{\lambda}^{scat} \sim 10 f_c$~\AA .  

Continuum fluorescence will enhance the line emission further via 
cascades from upper  
energy states. The importance of fluorescence versus scattering emission 
for a given line depends on the 
number of pumping transitions available to higher states, the 
optical depths in those transitions, 
the shape of the incident continuum, and the branching ratios 
(including escape probabilities) out of the upper states 
(see also Ferguson, Ferland \& Pradhan 1995). Experiments 
with multi-level atoms in CLOUDY indicate that the equivalent 
widths due to scattering {\it and} fluorescence are factors of 
$\sim$1.5 to $\sim$2 larger than the estimates above for scattering 
alone. 

If we add these illustrative scattering+fluorescence equivalent widths 
to the REWs in Figure 6, we would derive lower minimum covering 
factors and minimum column densities for the \ion{Ne}{8}~\lam 774 
emitting region by factors of a few compared to \S3.1.2. 
More importantly, the predicted contributions of \ion{N}{4}~\lam 765 
and \ion{O}{4}~\lam 789 to the measured $\sim$774~\AA\ lines would be 
several times larger than our estimates in \S3.1.1, because the 
theoretical ratio of \ion{N}{4}+\ion{O}{4} to \ion{C}{4} emission 
becomes larger. Therefore, \ion{Ne}{8}~\lam 774 might not dominate the 
average $\sim$774~\AA\ emission if there are large radial 
velocity dispersions that induce substantial continuum pumping. 
A thorough analysis of this 
possibility is beyond the scope of the present paper; we note simply 
that continuum pumping should also enhance the emission in other weak 
lines. Future studies might test the continuum pumping hypothesis by 
examining these weak lines and looking for correlations with the 
measured $\sim$774~\AA\ feature. For example, strong continuum pumping 
in \ion{N}{4}~\lam 765 and \ion{O}{4}~\lam 789 at $U\sim 0.1$ 
should be accompanied by strong pumped emission from 
\ion{C}{3}~\lam 977, \ion{N}{3}~\lam 991 and \ion{O}{3}~\lam 834. 
Similarly, strong pumping contributions to \ion{Ne}{8}~\lam 774 at 
$U\sim 10$ should 
come with highly pumped \ion{Mg}{10}~\lam 615 emission and an 
\ion{Mg}{10}/\ion{Ne}{8} flux ratio that is closer to unity than in 
Figure 6. Our limited CLOUDY experiments with $V_D = 1000$~\kms\ did 
not reveal any significant discrepancies with the data 
that would immediately rule out or require large pumping contributions. 

\subsection{Line Profiles and Redshifts}

\subsubsection{PKS 0355$-$483, Q~1435$-$015 and PG 1522+101}

The {\it HST} spectra of PKS~0355$-$483 and PG~1522+101 
(Figs. 1 and 4) have the strongest and most reliably 
measured emission lines near 774~\AA . 
Figures 7 and 8 show that their \Lya , \ion{C}{4}, \ion{N}{5} 
and \ion{O}{6} emission lines all have similar profiles.  
The only significant difference is that \ion{N}{5} and \ion{O}{6} 
are broader than \ion{C}{4} in PKS~0355$-$483. This small difference 
might be due to either the larger doublet separations in \ion{N}{5} and 
\ion{O}{6} (see Fig. 7) or small differences in the velocities in their 
emitting regions. In contrast, the $\sim$774~\AA\ features in both QSOs 
are dramatically broader than the other BELs (see also Table 2). 
Figure 9 shows similar results for Q~1435$-$015, although in this case 
comparisons with the $\sim$774~\AA\ line are more uncertain.  
Overall, we measure FWHMs in the $\sim$774~\AA\ feature that are roughly 2 
to 5 times larger than \ion{C}{4} and the other BELs in the same spectra 
(Table 2). 

Figures 10 and 11 show our attempts at fitting the $\sim$774~\AA\ 
features in PKS~0355$-$483 and PG~1522+101 
using the measured \ion{C}{4} profiles as templates. We constructed the 
templates by smoothing the \ion{C}{4} profiles without 
correcting for the \ion{C}{4} doublet separation (because it is 
relatively small, $\sim$500~\kms ). The top panels in Figures  
10 and 11 show the best fits assuming that \ion{Ne}{8} is the only  
contributor to the measured feature and both of the \ion{Ne}{8} doublet 
members (770 and 780~\AA ) have the same redshift and profile 
as \ion{C}{4}. The only freely varied parameters were the line strengths.  
The best fits were determined by $\chi^2$ minimization (with the 
narrow \Lya\ forest lines excluded from the minimization in PG~1522+101).    

The \ion{Ne}{8}-only fits show that the $\sim$774~\AA\ features are 
consistent with \ion{Ne}{8} emission at the \ion{C}{4} redshifts 
(see also Table 2 and Figs. 7--9 above). In PKS~0355$-$483 the 
redshifts of these lines are indistinguishable. 
In PG~1522+101 the purported \ion{Ne}{8} line is redshifted 
by up to $\sim$2000~\kms\ with respect to the \ion{C}{4} peak,  
but this shift might not be significant given the profile and 
centroid uncertainties in the broad $\sim$774~\AA\ feature. 
The much more significant result is that the \ion{C}{4} 
profiles are far too narrow to explain the $\sim$774~\AA\ features 
in terms of \ion{Ne}{8}. The lower two panels in Figures 10 and 11 plot 
the best fits with large contributions from 
\ion{N}{4}~\lam 765 and \ion{O}{4}~\lam 789. These fits assume 
that \ion{N}{4} and \ion{O}{4} also have the same redshift and 
profile as \ion{C}{4}, with the fluxes freely varied. 
The results using \ion{N}{4}~\lam 765 and 
\ion{O}{4}~\lam 789 alone (middle panels) account better for the overall 
width of the $\sim$774~\AA\ features but they do not match the profile 
shapes. Fits using \ion{N}{4} and \ion{O}{4} together with the two 
\ion{Ne}{8} doublet members (bottom panels) are much better, 
but we reiterate the conclusion from \S3.1; \ion{N}{4}~\lam 765 
and \ion{O}{4}~\lam 789 should be much weaker than they are portrayed 
in Figures 10 and 11, unless continuum pumping controls the line 
excitation.  

If the observed $\sim$774~\AA\ features are dominated by \ion{Ne}{8}, 
the velocity profiles must be substantially broader than \ion{C}{4} 
and other BELs in the same spectra. We can estimate the widths of 
the individual \ion{Ne}{8} doublet components by fitting the 
$\sim$774~\AA\ emisison in PKS~0355$-$483 and 
PG~1522+101 with one gaussian for each \ion{Ne}{8} doublet 
member. For these fits we require that the two gaussian components 
have the same redshift and velocity width,  
but the values of those parameters and the strengths 
of the lines are all freely varied. The results (shown in Figure 12) 
match the data reasonably well; the fit profiles are only slightly more 
centrally peaked than the observed \ion{Ne}{8} lines. 
In both QSOs, the observed 
lines are fit slightly better by a 1:1 doublet ratio, but 2:1 ratios 
are not ruled out. The centroids of the fits are within 0.3~\AA\ of 
the values listed in Table 2 and the individual components have 
FWHM~$\sim$~14,000 \kms\ in both QSOs. This velocity width is similar to 
the FWHMs measured directly from the profiles (Table 2). 

These results imply that the velocities in the \ion{Ne}{8} region are 
2--5 times larger than in the \ion{C}{4}-emitting gas. 
To understand how much of the \ion{Ne}{8} flux might come from the 
lower-velocity \ion{C}{4} gas, we performed additional 
fits to the PKS~0355$-$483 and PG~1522+101 profiles using (1) the 
\ion{C}{4} template at the \ion{C}{4} redshift 
(as in the top panels of Figs. 10 and 11) plus (2) a 
gaussian \ion{Ne}{8} doublet with freely-varied width and redshift
(as in Fig. 12). For both QSOs 
the $\chi^2$ minimization forced the \ion{C}{4} template contributions 
to zero, yielding fits identical to Fig. 12. Therefore, to first 
order, the measured profiles suggest emission from a high-velocity 
region with little or no contribution from the \ion{C}{4}-emitting gas.

\subsubsection{Mean Spectra}

We did not fit the \ion{Ne}{8}~\lam 774 profiles in the mean spectra 
because they might be broadened by uncertainties in the 
measured redshifts and by real redshift differences between \ion{Ne}{8} 
and the other lines.  Also, our selection requirement for measurable 
\ion{O}{6} emission might bias the Absnap and KP-sub samples toward more 
``peaky'' \ion{O}{6} line profiles. Nonetheless, we note that the 
mean \ion{Ne}{8} profiles are comparable to those in PKS~0355$-$483 and 
PG~1522+101 and more than twice as broad as both \ion{O}{6} and 
\Lya\ (Table 2). Although possibly spurious, this result is consistent 
with the individual QSOs discussed above. We also note that the 
$\sim$774~\AA\ centroid is slightly redshifted in the mean spectra 
(by $\sim$1000~\kms ) with respect to the \ion{O}{6} peak (Table 2); 
however, this velocity shift is probably smaller than the uncertainties. 

\section{Implications of Strong and Broad \protect{\ion{Ne}{8}~\lam 774} Emission}

The ionization parameters ($5\la U\la 30$) and covering factors 
($\ga$30\%) derived for the \ion{Ne}{8} emitting 
region (\S3.1.2) are both considerably larger than estimates 
based on low- to intermediate-ionization lines such as \Lya , 
\ion{Mg}{2}~\lam 2799, and \ion{C}{4}~\lam 1549. In particular, 
the latter lines indicate $U\la 0.2$ (\cite{dav79,kwa81,fer92,shi93,bal96}), 
while values as high as $U\sim 2$ have been inferred when 
\ion{O}{6}~\lam 1034  is also considered (\cite{ham92}). 
The covering factor of the low-ionization region is expected 
to be $\la$10\% based on both the ratio of observed to 
predicted \Lya\ equivalent widths and the infrequency of Lyman 
limit absorption edges at the emission redshift (see \cite{ost86}). 
The much larger covering factor of the \ion{Ne}{8} region 
implies that the high-ionization gas is optically thin in the 
\ion{H}{1} Lyman continuum (otherwise it would overproduce the \Lya\ 
emission). Optically thin, high-ionization clouds have been proposed 
before to explain the strength of the \ion{N}{5} and \ion{O}{6} 
lines (\cite{dav79}), and there is evidence from AGN variability studies 
that part of the \ion{C}{4}~\lam 1549 emission comes from optically 
thin gas (see \cite{shi95} and references therein). 
Our \ion{Ne}{8} results suggest that the highest 
ionization lines have the largest contributions from 
optically thin gas. George, Turner \& Netzer (1995) reached a similar 
conclusion based on the tentative detection of a large \ion{O}{7} 
line equivalent width in X-ray spectra of the Seyfert 1 galaxy NGC~3783. 

In spite of the low optical depths near the Lyman limit, the 
\ion{Ne}{8}-emitting gas has a substantial opacity to soft 
X-rays. Our calculations show that \ion{O}{7} and \ion{O}{8} are the 
dominant species of oxygen in the \ion{Ne}{8}-emitting region 
(see, for example, Fig. 3 in Hamann \etal 1995c). 
The minimum column density of $N_H \ga 10^{22}$~\cmsq\ 
(\S3.1.2) implies a minimum column in 
\ion{O}{7}+\ion{O}{8} of $N_{\rm O} \ga 10^{19}$~\cmsq . 
The resulting optical depth through the \ion{Ne}{8} gas is $\tau\ga 1$ 
at the \ion{O}{7} or \ion{O}{8} edges (based on photoionization 
cross-sections in \cite{ost89}). Therefore, the \ion{Ne}{8} component 
of the BELR would be a strong warm 
absorber, characterized by \ion{O}{7}+\ion{O}{8} edges 
at $\sim$0.8 keV, if it lies along our line of sight to the X-ray 
continuum source (see also \cite{net93,ham95a,shi95,net96}). 
Moreover, the range of physical conditions needed for strong \ion{Ne}{8} 
emission, $5\la U\la 30$ and $10^{22}\la N_H\la 3\times 10^{23}$~\cmsq\ 
(\S3.1.2), encompasses the specific conditions estimated for several 
strong warm absorbers (e.g. Mathur, Wilkes \& Aldcroft 1997, 
\cite{geo97,ota96,geo95,fio93}). The broad \ion{Ne}{8}~\lam 774 lines reported 
here clearly identify a BELR component with warm absorber properties. 

These results support models in which the  
warm absorbers are part of, or closely related to, the 
BELR (e.g. \cite{rey95,net96} and references therein). Other 
studies suggest that warm absorbers are also closely related to the 
blueshifted absorption lines observed in some QSOs and Seyfert 
galaxies (\cite{mat94a,mat94b,mat95,shi97}). These absorption lines 
form in outflows with line-of-sight velocities ranging from near 
zero to $\ga$20,000~\kms\ (e.g. in the broad absorption line [BAL] 
systems; \cite{fol88,ulr88,wey91}). 
There is no direct evidence for outflow in the 
\ion{Ne}{8}~\lam 774 emission-line data, but the 
ultra-broad line widths reported 
here, with FWHM~$\sim$~14,500~\kms\ in two well-measured cases, 
are consistent with high velocities in the warm absorber gas. 
The large \ion{Ne}{8} FWHMs are also consistent with 
emission from the ``very broad line region'' (VBLR), 
which was identified by Ferland, Korista 
\& Peterson (1990) from the non-varying \Ha\ line wings 
in a Seyfert 1 galaxy. Ferland \etal (1990) present independent arguments 
that the VBLR is highly ionized, optically thin at the Lyman limit 
and the likely source of X-ray warm absorption (see also \cite{mar93}).

The large covering factors and warm absorber-like properties 
of the \ion{Ne}{8} emitting regions suggest 
that warm absorbers should be common in QSOs. 
The most recent X-ray surveys indicate that 
warm absorbers are common in Seyfert 1 nuclei (with a $\ga$50\% 
detection rate; \cite{tur89,tur93b,nan94,rey97,geo97}) 
but {\it rare} in QSOs ($\sim$5\% detection rate; \cite{lao97b}). 
The rarity of warm absorbers in QSOs might be reconciled with 
the \ion{Ne}{8} emission-line data if (1) QSO ionizing spectra are 
harder than the Mathews \& Ferland (1987) continuum, requiring smaller 
\ion{Ne}{8} covering factors (\S3.1.2), (2) large radial velocity 
dispersions ($\ga$1000~\kms ) in the BELR enhance the emission in 
\ion{Ne}{8}~\lam 774 and other weak lines via continuum pumping 
(\S3.1.3), or (3) QSOs are observed 
over limited viewing angles that usually 
avoid \ion{Ne}{8} gas along our line-of-sight to the X-ray 
continuum source. 
Viewing angles that avoid the BELR might occur naturally in the 
recent disk-wind models (\cite{mur95a,mur95b}). In these models, 
all of the BELs form close to the accretion disk and   
the highest ionization lines 
form at the smallest disk radii.  Murray (1996) pointed out that the 
exceptionally broad \ion{Ne}{8}~\lam 774 profiles 
might be due to enhanced rotational broadening 
associated with the inner disk. However, that interpretation 
might conflict with our estimates of large \ion{Ne}{8} 
covering factors, which require 
substantial emission from regions away from the disk plane. 
Better data and more specific comparisons with the models are needed 
before drawing firm conclusions. 

The large \ion{Ne}{8}~\lam 774 line widths compared to 
\ion{C}{4}~\lam 1549 and especially \ion{O}{6}~\lam 1034 might provide 
strong constraints on models of the emitting regions in general. 
The calculations 
in Figure 6 indicate that the gas producing the ultra-broad 
\ion{Ne}{8} line should also produce considerable \ion{O}{6} emission. 
This ultra-broad component to the \ion{O}{6} lines might be hard to 
hide in the measured profiles, particularly for sources like 
PKS~0355$-$483 where any high-velocity \ion{O}{6} emission must be 
weak compared to \ion{Ne}{8}~\lam 774 (Figs. 1 and 7). 
However, we can lower the predicted \ion{O}{6}/\ion{Ne}{8} flux ratio by 
adopting large values of $U$ and/or lower $N_H$ compared to that used 
in Figure 6. (Adding continuum pumping via large turbulent velocities 
[\S3.1.3] does not significantly alter the \ion{O}{6}/\ion{Ne}{8} 
flux ratio.) Limited experiments with CLOUDY indicate that 
\ion{O}{6}~\lam 1034 can be $\sim$2 or more times weaker than 
\ion{Ne}{8}~\lam 774 if $U\geq 20$ and $N_H \leq 3\times 10^{22}$~\cmsq . 
Larger gas densities and/or different ionizing continuum shapes  
can further reduce the theoretical \ion{O}{6}/\ion{Ne}{8} flux ratio 
(see also \cite{kor97}). 
Alternatively, or perhaps in addition, the measured $\sim$774~\AA\ lines 
might be broadened by blending with 
\ion{N}{4}~\lam 765 and \ion{O}{4}~\lam 789 if their emission is enhanced 
by continuum pumping (\S\S3.1.3 and 3.2.1). In that case, the true 
\ion{Ne}{8} line widths could be comparable to the other BELs 
and there would be no need for an ultra-high velocity 
\ion{Ne}{8} emitting region. 

\section{Summary}

The mean spectrum of an unbiased sample (KP-full) of 11 radio-loud and 
radio-quiet QSOs clearly shows broad-line emission at $\sim$774~\AA\ 
with an average REW of $\sim$5.9~\AA\ and an average 
flux equal to $\sim$42\% of \ion{O}{6}~\lam 1034. 
This result supports our previous conclusion (based on a smaller 
sample) that $\sim$774~\AA\ emission is common in QSOs. 
New {\it HST}-FOS spectra of individual QSOs reveal a 
strong and broad $\sim$774~\AA\ line in PG~1522+101 (also \cite{coh94}), 
and they support the prior claims of strong $\sim$774~\AA\  
lines in PKS~0355$-$483 and Q~1435$-$015 (\cite{ham95b}). 
However, the new spectra contradict the report (based on old {\it IUE} data) 
that the $\sim$774~\AA\ line is strong in PG~1148+549 (\cite{ham95a}). 

The centroids of the observed $\sim$774~\AA\ features support  
\ion{Ne}{8} as the single most likely identification, 
but the velocity widths are 
significantly larger than \ion{O}{6}~\lam 1034 and 
all other BELs in two well-measured 
QSOs (PKS~0355$-$483 and PG~1522+101) and probably in the mean 
spectra. The $\sim$774~\AA\ line widths, ranging from 
FWHM~$\approx$~11,000 to 15,000~\kms , are $\sim$2 to 5 times larger 
than the lower-ionization BELs. These ultra-broad profiles are 
consistent with significant contributions from \ion{N}{4}~\lam 765 and 
\ion{O}{4}~\lam 789 emission. However, photoionization calculations 
coupled with measurements of \ion{C}{4}~\lam 1549 constrain the 
\ion{N}{4} and \ion{O}{4} line strengths to be at least several times 
weaker than the average $\sim$774~\AA\ emission (\S3.1.1). We conclude 
that exceptionally broad \ion{Ne}{8}~\lam 774 
dominates the average feature and is the only significant contributor 
when this line is strong compared to \ion{C}{4}~\lam 1549. The only 
way to avoid this conclusion is to invoke large line-of-sight velocity 
dispersions ($\ga$1000~\kms ) in the BELR, so that the resonant absorption 
of continuum photons greatly enhances the \ion{N}{4} and \ion{O}{4} line 
strengths relative to both the continuum and \ion{C}{4}~\lam 1549 (\S3.1.3). 
If \ion{Ne}{8} dominates the $\sim$774~\AA\ emission, fits to the line 
profiles indicate that most of the flux comes from a distinct 
high-velocity region, with little or no contribution 
from the low-velocity gas responsible for  
\ion{C}{4} and the other lower-ionization lines (\S3.2.1).

The photoionization calculations also indicate that the \ion{Ne}{8} 
emitting regions have ionization parameters $5\la U\la 30$, total 
column densities $10^{22}\la N_H\la 3\times 10^{23}$~\cmsq , and covering 
factors $\ga$30\% of 4$\pi$~steradians (for solar abundances and a standard 
ionizing spectrum; \S3.1.2). These results are uncertain by factors of a few 
because of the poorly known shape of the far-UV spectrum and the 
possible large contributions to the line fluxes from continuum pumping. 
Nonetheless, the \ion{Ne}{8} component of the BELR must be considerably
 more ionized and probably more extensive than the regions producing 
lines like \Lya , \ion{Mg}{2}~\lam 2799, and \ion{C}{4}~\lam 1549. 
The physical conditions in the \ion{Ne}{8} emitting clouds are 
characteristic of the highly-ionized ``warm'' absorbers observed in 
soft X-rays, suggesting 
a close relationship between the BELR and the warm absorber gas. 
The \ion{Ne}{8} emitting regions would themselves produce warm 
absorption (e.g. bound-free edges of \ion{O}{7}+\ion{O}{8} at 
$\sim$0.8 keV) if they lie along our line of sight to the 
X-ray continuum source.

\acknowledgments
We are grateful to G. J. Ferland for providing the CLOUDY software and 
R. Weymann for the reduced {\it HST} Key Project spectra. We also 
thank L. Zuo and D. Tytler for help with the Absnap data, 
G. J. Ferland, K. Korista and A. Laor for useful 
discussions, and R. Lyons for comments on this manuscript and 
help with the data processing. Finally, we thank an anonymous referee 
for helpful suggestions. 
This work was supported by NASA grants NAG~5-1630 and 
NAG~5-3234, and by the Space Telescope Science Institute through 
the archival grant AR-5292-93A and the Guest Observer grant 
GO-6103-94A.

\clearpage
{
\begin{deluxetable}{lcccrr}
\tablecaption{Log of Observations}
\tablecolumns{6}
\tablewidth{0pt}
\tablehead{
\colhead{Date}& \colhead{Tel.}&  \colhead{Instr./Setup} & 
\colhead{$\lambda_{obs}$\tablenotemark{a}}& \colhead{$R$}& 
\colhead{Exp.\tablenotemark{a}} \\
\cline{1-6}
\noalign{\vskip 7pt}
\multicolumn{6}{c}{PKS 0355$-$483 ($z_e = 1.005$)}
}
\startdata
12 Feb. 1996& {\it HST}& FOS/BL--G160L& 1140--2508& 250& 5440\nl
12 Feb. 1996& {\it HST}& FOS/BL--G270H& 2221--3301& 1300& 1090\nl
\cutinhead{PG 1148+549 ($z_e = 0.969$)}
15 Sep. 1995& {\it HST}& FOS/BL--G130H& 1140--1606& 1300& 2860\nl
15 Sep. 1995& {\it HST}& FOS/BL--G190H& 1573--2330& 1300& 750\nl
15 Sep. 1995& {\it HST}& FOS/BL--G270H& 2221--3301& 1300& 350\nl
\cutinhead{Q 1435$-$015 ($z_e = 1.310$)}
21 Jan. 1996& {\it HST}& FOS/BL--G160L& 1140--2508& 250& 2960\nl
21 Jan. 1996& {\it HST}& FOS/BL--G270H& 2221--3301& 1300& 720\nl
30 Jul. 1995& Lick 3m& KAST--grismIII& 3100--4415& 1400& 3000\nl 
\cutinhead{PG 1522+101 ($z_e = 1.318$)}
14 Mar. 1992& {\it HST}& FOS/BL--G160L& 1140--2508& 250& 5635\nl 
14 Mar. 1992& {\it HST}& FOS/RD--G190H& 1573--2330& 1300& 9487\nl 
14 Mar. 1992& {\it HST}& FOS/RD--G270H& 2221--3301& 1300& 2638\nl 
27 Mar. 1992& Lick 3m& KAST--grismIII& 3250--4600& 1400& 3600\nl 
28 Mar. 1992& Lick 3m& KAST--grismIII& 3150--4500& 1400& 3600\nl 
\tablenotetext{a}{$\lambda_{obs}$ is in \AA\ and Exp. is in seconds.}\nl
\enddata
\end{deluxetable}
\begin{deluxetable}{lccccc} 
\tablewidth{0pt}
\tablecaption{Emission Line Measurements\tablenotemark{a}}
\tablecolumns{6}
\tablehead{
\colhead{Line} &
\colhead{$\lambda_{rest}$} &
\colhead{REW} &
\colhead{Flux\tablenotemark{b}} &
\colhead{$\Delta\lambda$} &
\colhead{FWHM} \nl
\colhead{} &
\colhead{(\AA )} &
\colhead{(\AA )} &
\colhead{} &
\colhead{(\AA )} &
\colhead{(\kms )}
}
\startdata
\cutinhead{Total Mean (16 spectra)}
\ion{Ne}{8}~$\lambda$774 \ \ \ & 776.3 & 7.5 & 0.51 & 745--804& 15,300\nl
\ion{N}{3} + \ion{C}{3}& 982.0& 5.0& 0.31& 963--997& \nodata\nl
\ion{O}{6}~$\lambda$1034& 1032.0 & 16.0 & 1.00 & \hskip 0.5em 
997--1057& 6200\nl
\ \ \ \ \ ???& 1070.9 & 3.5 & 0.22 & 1057--1090& \nodata\nl
Ly$\alpha$ + \ion{N}{5} + \ion{Si}{2}& 1221.3& 87.0& 5.2& 1160--1290& 
\nodata\nl
%
\cutinhead{KP-full Mean (11 spectra)}
\ion{Ne}{8}~$\lambda$774 \ \ \ & 776.3 & 5.9 & 0.42 & 745--804& 14,800\nl
\ion{N}{3} + \ion{C}{3}& 982.0& 4.6& 0.30& 963--997& \nodata\nl
\ion{O}{6}~$\lambda$1034& 1030.8 & 15.3 & 1.00 & \hskip 0.5em 
997--1057& 6500\nl
\ \ \ \ \ ???& 1070.5 & 2.8 & 0.18 & 1057--1090& \nodata\nl
Ly$\alpha$ + \ion{N}{5} + \ion{Si}{2}& 1220.8& 85.9& 5.3& 1160--1290& 
\nodata\nl
%
\cutinhead{Absnap Mean (5 spectra)}
\ion{Ne}{8}~$\lambda$774 \ \ \ & 776.5 & 10.5 & 0.85 & 748--803& 11,200\nl
\ion{N}{3} + \ion{C}{3}& 982.8& 6.0& 0.44& 961--1001& \nodata\nl
\ion{O}{6}~$\lambda$1034& 1033.9 & 13.5 & 1.00 & \hskip 0.5em 
1001--1057& 5700\nl
\ \ \ \ \ ???& 1073.2 & 4.3 & 0.32 & 1057--1095& \nodata\nl
Ly$\alpha$ + \ion{N}{5} + \ion{Si}{2}& 1221.0& 84.2& 5.9& 1160--1290& 
\nodata\nl
%
\cutinhead{PKS 0355$-$483 ($z_e = 1.005$)}
\ion{Ne}{8}~$\lambda$774& 773.9& 13.8& 9.3& 748--800& 14,100\nl
\ion{N}{3} + \ion{C}{3}& 987.9& 6.5& 4.0& 967--1004& \nodata\nl
\ion{O}{6}~$\lambda$1034& 1034.7& 10.6& 6.4& 1017--1049& 3750\nl
Ly$\alpha$ + \ion{N}{5} + \ion{Si}{2}& 1225.9& 67.5& 36.5& 1193--1283& 
\nodata\nl
\ion{Si}{4} + \ion{O}{4}]& 1402.8& 4.1& 1.9& 1377--1429& \nodata\nl
\ion{C}{4}~$\lambda$1549& 1551.6& 8.3& 3.4& 1536--1575& 2500\nl
%
\cutinhead{PG 1148+549 ($z_e = 0.969$)}
\ion{Ne}{8}~$\lambda$774& 772.4& 4.0\rlap{:}& 3.4\rlap{:}& 743--801& \nodata\nl
\ion{N}{3} + \ion{C}{3}& 983.1& 4.9\rlap{:}& 3.5\rlap{:}& 962--1000& \nodata\nl
\ion{O}{6}~$\lambda$1034& 1025.5& 14.2& 9.9& 1000--1047& 8500\nl
Ly$\alpha$ + \ion{N}{5} + \ion{Si}{2}& 1219.7& 89.5& 53.2& 1155--1275& 
\nodata\nl
\ion{Si}{4} + \ion{O}{4}]& 1398.7& 10.0& 4.7& 1373--1418& \nodata\nl
\ion{C}{4}~$\lambda$1549& 1540.8& 35.1& 17.8& 1488--1586& 7400\nl
\tablebreak
\cutinhead{Q 1435$-$015 ($z_e = 1.310$)}
\ion{Ne}{8}~$\lambda$774& 775.6& 5.4& 4.1& 746--805& 11,000\rlap{:}\nl
\ion{N}{3} + \ion{C}{3}& 977.6& 6.6& 4.6& 951--998& \nodata\nl
\ion{O}{6}~$\lambda$1034& 1031.1& 11.2& 7.7& 1003--1059& 6800\rlap{:}\nl
Ly$\alpha$ + \ion{N}{5} + \ion{Si}{2}& 1219.3& 62.7& 41.1& 1165--1283& 
\nodata\nl
\ion{Si}{4} + \ion{O}{4}]& 1397.3& 4.2& 2.3& 1370--1425& \nodata\nl
\ion{C}{4}~$\lambda$1549& 1546.5& 26.0& 13.7& 1476--1597& 4200\nl
\cutinhead{PG 1522+101 ($z_e = 1.318$)}
\ion{Ne}{8}~$\lambda$774& 776.1& 13.5& 12.1& 739--808& 15,000\nl
\ion{O}{6}~$\lambda$1034& 1032.2& 17.0& 16.9& \hskip 0.5em 998--1063& 
7200\nl
Ly$\alpha$ + \ion{N}{5} + \ion{Si}{2}& 1224.2& 63.1& 58.5& 1166--1287& 
\nodata\nl
\ion{Si}{4} + \ion{O}{4}]& 1406.6& 10.1& 8.3& 1368--1439& \nodata\nl
\ion{C}{4}~$\lambda$1549& 1544.8& 27.1& 20.8& 1489--1592& 6500\nl
\enddata
\tablenotetext{a}{Wavelengths and REWs are relative to the redshifts 
listed.}
\tablenotetext{b}{Fluxes are relative to \ion{O}{6}~\lam 1034 
for the mean spectra and as observed with units 10$^{-14}$ 
ergs s$^{-1}$ cm$^{-2}$ for the individual QSOs.}
\end{deluxetable}
}
\clearpage

\clearpage

\centerline{\bf Figure Captions}
\bigskip  
\figcaption[dum]{{\it HST}-FOS spectra of PKS 0355$-$483. The ``Absnap'' 
spectrum is from Hamann \etal (1995b). The ``new {\it HST}'' spectrum 
(this paper) is shifted vertically by adding 3.5 in these $F_{\lambda}$ 
flux units, 10$^{-15}$ \flam . The narrow peak in \Lya\ is due at least 
partly to geocoronal \Lya\ emission in second order at $\sim$2435~\AA . 
Both spectra are smoothed twice with a 3-pixel-wide boxcar function. 
The dotted curve is a fit to the continuum. Various possible emission lines 
are labeled at the redshift given in the tables; not all of these lines are 
present. See \S2.1. \label{fig1}}

\figcaption[dum]{{\it HST}-FOS and {\it IUE} spectra of PG~1148+549. The 
``{\it HST}'' and ``new {\it IUE}'' spectra are shifted up by adding 
22.0 and 11.0, respectively.  The {\it IUE} spectra are smoothed twice 
and the {\it HST} spectrum 4 times with a 3-pixel-wide boxcar function. 
The spike at 1663~\AA\ in the new {\it IUE} spectrum is a camera 
artifact. The old and new {\it IUE} spectra are different reductions 
of the same data. See Figure 1 and \S2.1. \label{fig2}}

\figcaption[dum]{{\it HST}-FOS spectra of Q~1435$-$015. The ``new 
{\it HST} + Lick'' spectrum is shifted up by adding 2.5, and both 
{\it HST} spectra are smoothed twice 
with a 3-pixel-wide boxcar function. See Figure 1 and \S2.1. 
\label{fig3}}

\figcaption[dum]{{\it HST}-FOS and Lick Observatory spectra of 
PG~1522+101. The G160L spectrum is shifted up by adding 
3.3. The high-resolution spectrum is smoothed by 3 applications 
of a 3-pixel-wide boxcar function. Weak Lyman limit absorption at 
$\sim$1980~\AA\ affects the continuum shape. See Figure 1 and \S2.1. 
\label{fig4}}

\figcaption[dum]{Mean {\it HST}-FOS spectra of QSOs in the Absnap, 
KP-sub, KP-full and Total samples (see \S2.2). 
The lines are labeled at their rest wavelengths. 
The continua are normalized to unity 
but shifted vertically by adding 1.0 to the KP-sub mean, 
2.0 to KP-full and 3.0 to the Total. 
The dotted curves show the estimated continuum levels. 
The histogram at the bottom shows the number of spectra 
contributing to the Total mean at each wavelength. 
\label{fig5}}

\figcaption[dum]{Predicted line equivalent widths ($W_{\lambda}$ in~\AA ) 
for photoionized emitting regions that completely cover the central 
continuum source (over 4$\pi$~steradians). Each value of the ionization 
parameter represents a different region/calculation. The equivalent 
widths can be converted to relative fluxes by scaling by the 
continuum slope across these wavelengths, 
$F_{\lambda}\propto \lambda^{-1.5}$. See \S3.1.\label{fig6}}

\figcaption[dum]{Line profiles in PKS 0355$-$483 (solid curves, as labeled) 
are compared to \protect\ion{C}{4}~\lam 1549 (dotted curve in all panels). 
Zero velocity is defined by the emission redshift given in Table 1, 
assuming rest wavelengths weighted 2:1 for the doublets 
\protect\ion{C}{4}~\lam\lam 1548,1551, \protect\ion{N}{5}~\lam\lam 1239,1243 
and \protect\ion{O}{6}~\lam\lam 1032,1038.
Open brackets mark the positions of the 
individual doublet members. The \protect\ion{C}{4} doublet separation 
(not shown) is 500 \kms . The velocity scale in the bottom panel applies 
to \Lya\ 1215.67~\AA . The spectra are  normalized to unity 
in the continuum and smoothed by 1 application of a 
3-pixel-wide boxcar function. See \S3.2.1. \label{fig7}}

\figcaption[dum]{Same as Figure 7 but for PG~1522+101. \label{fig8}}

\figcaption[dum]{Same as Figure 7 but for Q~1435$-$015. \label{fig9}}

\figcaption[dum]{\protect\ion{Ne}{8}~\lam 774 line profile (bold curve in 
all panels) measured in PKS~0355$-$483 is compared to various fits 
(thin curves) that use the \protect\ion{C}{4}~\lam 1549 profile as a 
template. The lines contributing to the fits are indicated in each panel, 
with tic marks across the bottom showing their centroid positions.  
See Figure 7 and \S3.2.1. \label{fig10}}

\figcaption[dum]{Same as Figure 9 but for PG~1522+101. \label{fig11}}

\figcaption[dum]{Measured \protect\ion{Ne}{8}~\lam 774 line profiles 
(bold curves) in PKS~0355$-$483 and PG~1522+101 are compared to 
gaussian fits (one gaussian per doublet component). The thin 
solid line shows the overall fit, while the dotted lines show 
the individual components. The component centroids are shown by 
tic marks at the bottom of each plot. See \S3.2.1. \label{fig12}}

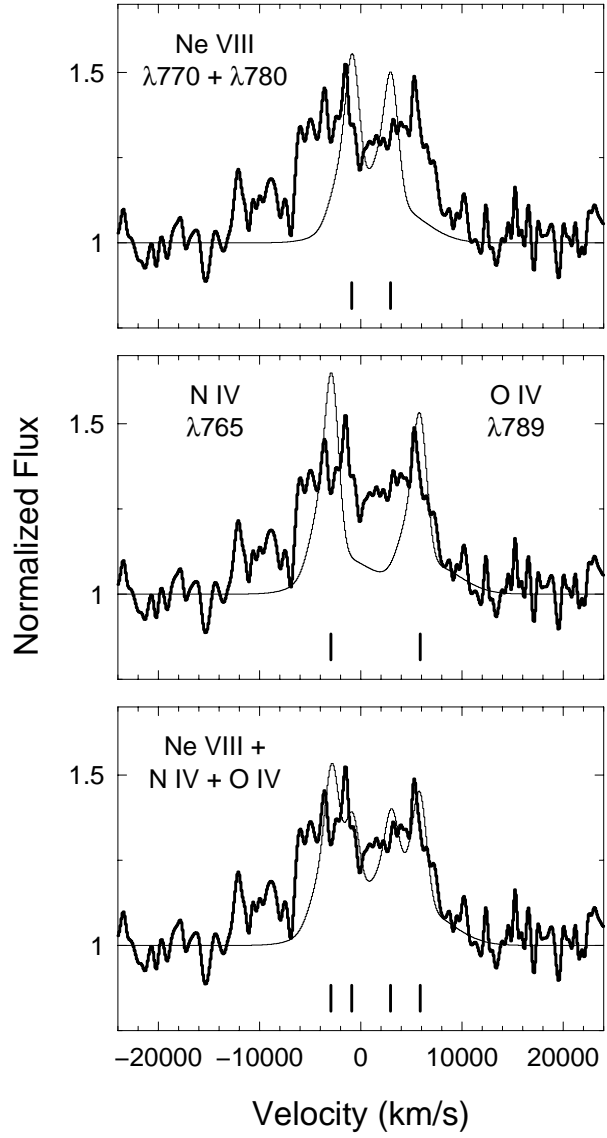
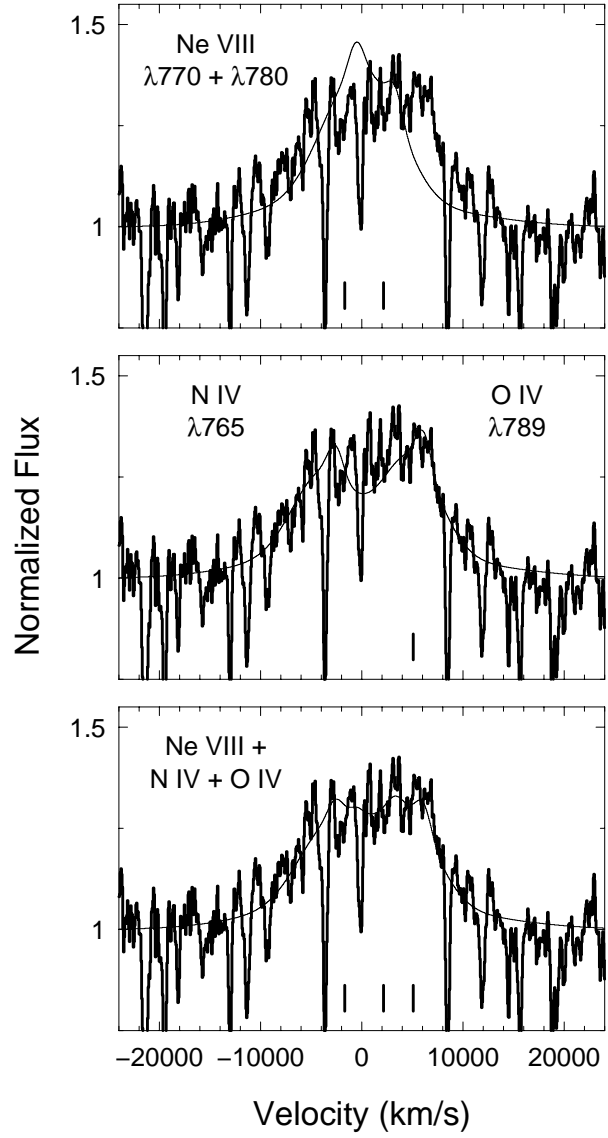
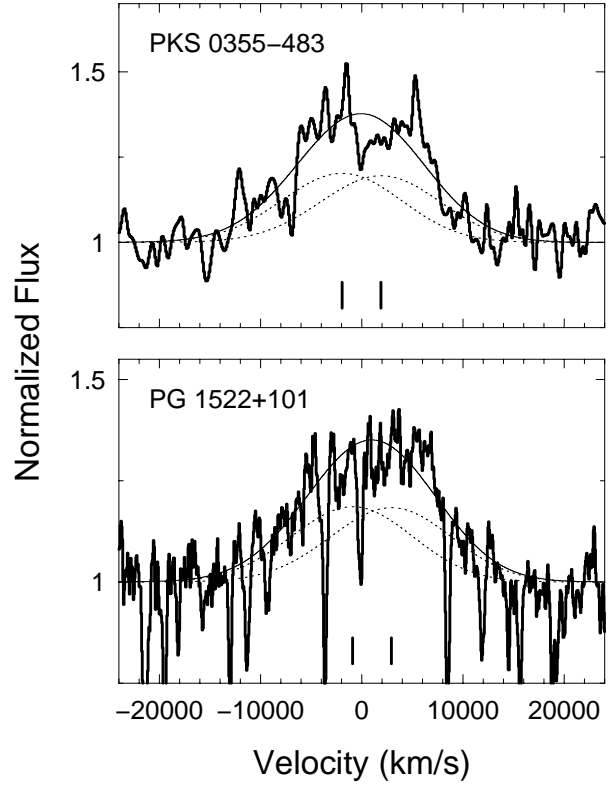
\end{document}